\documentclass{sig-alternate}
\usepackage[small,it]{caption}
\usepackage{amsmath,amsfonts,amssymb}
\usepackage{graphicx}
\usepackage{epstopdf}
\usepackage{paralist}
\usepackage{tabularx}
\usepackage{multirow}
\usepackage{color}

\newcommand{\cC}{{\cal C}}

\newcommand{\RR}{\mathbb{R}}

\newcommand{\cei}[1]{\lceil #1 \rceil}

\newcommand{\Sec}[1]{\hyperref[sec:#1]{\S\ref*{sec:#1}}} 
\newcommand{\App}[1]{\hyperref[sec:#1]{Appendix~\ref*{sec:#1}}} 
\newcommand{\Fig}[1]{\hyperref[fig:#1]{Fig.\,\ref*{fig:#1}}} 
\newcommand{\Thm}[1]{\hyperref[thm:#1]{Thm.\,\ref*{thm:#1}}} 
\newcommand{\Lem}[1]{\hyperref[lem:#1]{Lem.\,\ref*{lem:#1}}} 
\newcommand{\Prop}[1]{\hyperref[prop:#1]{Prop.~\ref*{prop:#1}}} 
\newcommand{\Cor}[1]{\hyperref[cor:#1]{Cor.~\ref*{cor:#1}}} 
\newcommand{\Def}[1]{\hyperref[def:#1]{Defn.~\ref*{def:#1}}} 
\newcommand{\Ex}[1]{\hyperref[ex:#1]{Ex.~\ref*{ex:#1}}} 
\newcommand{\Clm}[1]{\hyperref[clm:#1]{Claim~\ref*{clm:#1}}} 
\newcommand{\Step}[1]{\hyperref[step:#1]{Step~\ref*{step:#1}}} 

\newcommand{\cema}{\cC}
\newcommand{\cmetric}{C}
\newcommand{\pmetric}{P}
\newcommand{\pt}{{p}}
\newcommand{\samp}{k}
\newcommand{\thresh}{\tau}

\newcommand{\be}{\begin{equation}}
\newcommand{\ee}{\end{equation}}
\newcommand{\bea}{\begin{eqnarray}}
\newcommand{\eea}{\end{eqnarray}}

\begin{document}
\title{ Enabling adaptive scientific workflows via trigger detection\thanks{This work was funded by the Laboratory Directed
Research and Development (LDRD) program of  Sandia National Laboratories. Sandia
National Laboratories is a multi-program laboratory managed and operated by
Sandia Corporation, a wholly owned subsidiary of Lockheed Martin Corporation,
for the U.S. Department of Energy's National Nuclear Security Administration
under contract DE-AC04-94AL85000.}}
\numberofauthors{1} 
\author{
\alignauthor Maher Salloum \quad  Janine C. Bennett  \quad Ali Pinar \\
Ankit Bhagatwala  \quad Jacqueline H. Chen\\
       \affaddr{Sandia National Laboratories,  Livermore, CA }\\
       \email{\{mnsallo, jcbenne, apinar, abhagat, jhcehn\} @sandia.gov}
}
\maketitle
\begin{abstract}
Next generation architectures necessitate a shift away from traditional workflows in
which the simulation state is saved at prescribed frequencies for
post-processing analysis. While the need to shift to in~situ workflows has been 
acknowledged for some time, much of the current research is focused on
\emph{static workflows}, where the analysis that would have been done as a
post-process is performed concurrently with the simulation at user-prescribed
frequencies.  Recently, research efforts are striving to enable \emph{adaptive workflows}, 
in which the frequency, composition, and execution of  computational and data
manipulation steps dynamically depend on the state of the simulation.  Adapting
the workflow to the state of simulation in such a data-driven fashion puts
extremely strict efficiency requirements on the analysis capabilities that are used  
to identify the transitions  in the  workflow. 
In this paper we build upon earlier work on trigger detection using sublinear
techniques to drive adaptive workflows.  Here we propose a methodology  to
detect the time when sudden heat release occurs in
simulations of turbulent combustion.  Our proposed method provides an alternative metric that can be used 
along with our former metric to increase the robustness 
of trigger detection.  We show the effectiveness of our metric empirically for
predicting heat release for two use cases. 
\end{abstract}

\section{Introduction}
Concurrent analysis frameworks have been developed to process  raw simulation output as it is computed, decoupling the analysis from I/O. 
Operations sharing primary resources of the simulation are considered \emph{in~situ}, while \emph{in~transit} processing involves asynchronous data transfers to secondary resources. 
Both in~situ~\cite{Yu2010,visit:2011, paraview:ldav11}  and
in~transit~\cite{glean:ldav11, JITStaging, Bennett:2012} workflows perform
analyses as the simulation is run and produce results, which are typically much
smaller than the raw data, mitigating the effects of limited disk bandwidth and
capacity.  Concurrent analyses are often performed at prespecified frequencies, which is  viable  for analyses that are not too expensive -- in terms of runtime
(with respect to a simulation time step), memory footprint, and output size.  However, for many  other analyses 
that are resource intense, prescribed frequencies will not suffice because the 
scientific phenomenon being simulated typically does not behave linearly (e.g., combustion, climate,
astrophysics).  When the prescribed I/O or analysis frequency  is high enough to capture the events of interest, the costs incurred are 
too great. On the other hand, a cost-effective, lower analysis-frequency  may miss the scientific events that simulation is intended to capture.

An alternative approach is to perform expensive analyses (denoted $A_e$) and I/O in an
adaptive fashion, driven by the data itself.   A domain-agnostic approach is
presented in~\cite{nouanesengsy2014adr,
modelPaper}, based on entropy of information in the data.   
Our previous work in~\cite{cema-insitu} provides a framework for 
making data-driven control-flow decisions that can leverage the scientists'
intuitions, even when the algorithms to capture those
intuitions would otherwise be too expensive to compute. 
Our methodology involves a user-defined \emph{indicator} function that is computed and measured in~situ 
at a relatively regular and high-frequency (i.e., greater than the frequency
with which the I/O or $A_e$ would be prescribed).  Along with the indicator, the application scientist
defines an associated \emph{trigger}, a function that returns a boolean value
indicating that the indicator has met some property, for example a threshold
value.  While our  methodology is intuitive and conceptually quite simple,
the challenges lie in defining indicators and triggers that capture the appropriate
scientific information, while remaining cost efficient in terms of runtime, memory footprint, and I/O
requirements, so that they can be deployed at the high frequency that is required.  

In~\cite{cema-insitu} we show 
that chemical explosive mode analysis,  denoted CEMA, can be used to devise a 
noise-tolerant indicator for  heat release (the quantity of interest that the
scientists would like to capture), thus making it a good candidate to drive adaptive workflows. However, 
exhaustive computation of CEMA values dominates the  simulation time.  To overcome this 
bottleneck, we proposed a quantile sampling approach with provable
error/confidence bounds. These bounds depend only on  
 the number of samples and  are independent of the problem size.  We also designed an indicator, referred to as 
 the \emph{\pmetric-indicator},  based on the quantile sampling approach.  Our experiments on homogeneous charge compression ignition
(HCCI)  and reactivity controlled compression ignition (RCCI) simulations showed
 that the proposed method can detect rapid increases in heat release \emph{and} is computational efficient.

In this paper, we propose an alternative indicator, referred to as the
\emph{\cmetric-indicator}, and associated trigger function. 
The proposed technique is based on the coefficient of variation and similar in 
essence to our earlier technique that used quantiles, as it  tries to detect 
shrinking in the range of upper percentiles in the distribution of CEMA values.  
The coefficient of variation among the top quantiles decreases as the  range of the
values shrink and  can serve as a detector for the subsequent heat release
that  the scientists wish to capture. 
The new technique based on  coefficient of variation is not  proposed as an
alternative to replace our former metric. Rather, we  believe that a set of trigger detection mechanisms, 
when used collectively,  can provide more robust detections, as  collectively
they will be  less prone to false positives. 
Our experiments on HCCI  and RCCI simulations 
show that the proposed method can  efficiently detect rapid increases in heat release.

\section{Combustion as a use case for \\trigger detection}
\begin{figure}[t]
\centering
\includegraphics[width=1.0\columnwidth]{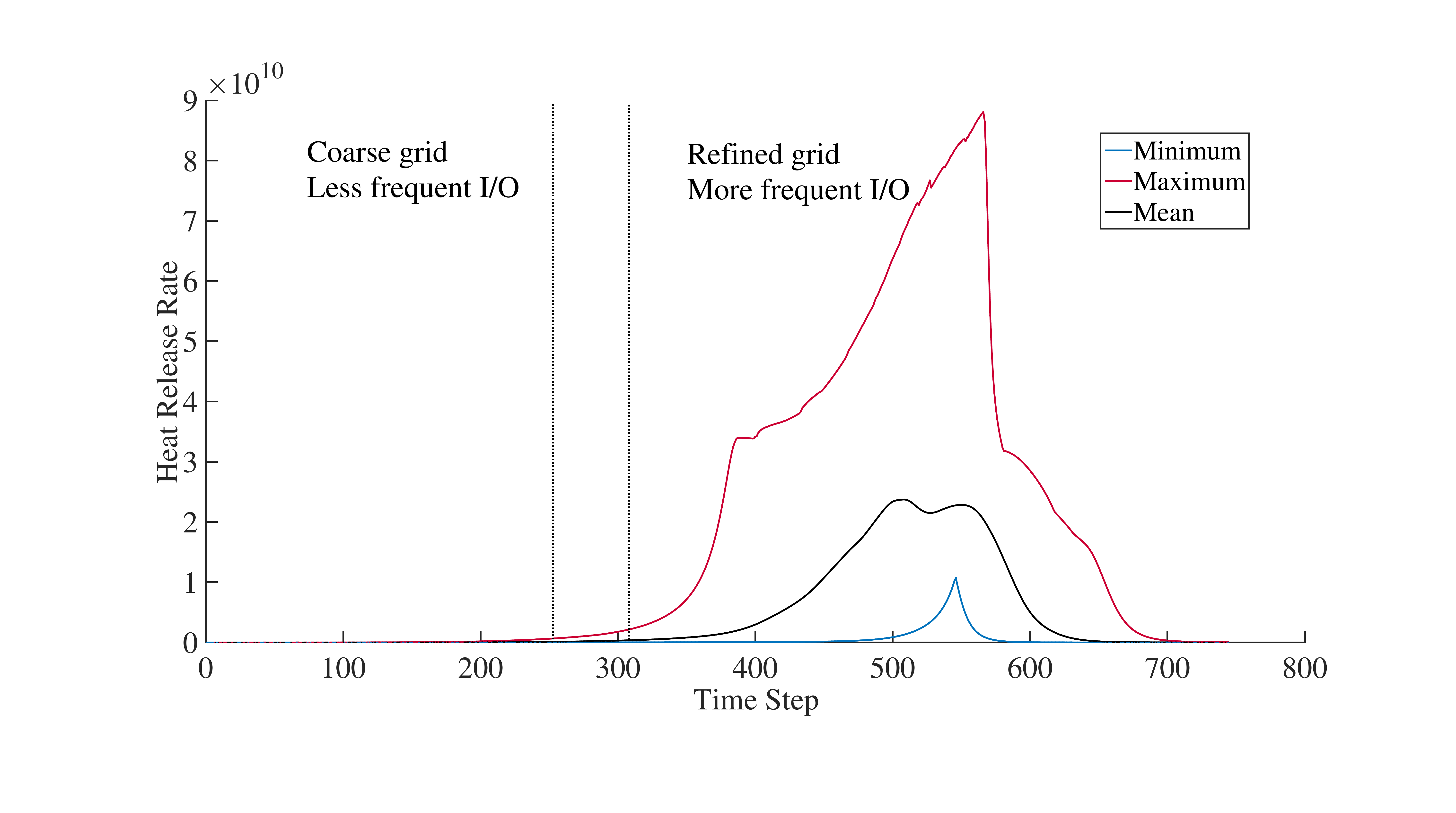}
\caption{\label{fig:heat_release} 
  The minimum (blue), maximum (red) and mean (black) heat release values
  for each time step in the simulation. Early in the simulation,
  we want to run at a coarser grid  and save data less frequently.
  When heat release occurs, we want to the finest grid and save the data as frequently as possible.  The vertical lines define a range of time steps within which we would like to make
  this transition, as identified by a domain expert.
}
\end{figure}

We demonstrate our approach applied to a combustion
use case, using S3D~\cite{chen09}, a direct numerical simulation of 
combustion in turbulence.  
The combustion simulations in our use case pertain to a class of internal combustion
(IC) engine concept. 
We are interested in two specific techniques called homogeneous-charge compression
ignition (HCCI) \cite{bhagatwala1} and reactivity-controlled compression
ignition (RCCI) \cite{kokjohn,bhagatwala2}. In both cases, heat release
starts in the form of small kernels at arbitrary locations.  Eventually, multiple kernels 
ignite as the overall heat release reaches a global maximum and subsequently 
declines. Since these simulations are computation and storage-intense, 
we want to run the simulation at a coarser grid resolution and save data less 
frequently during the early build-up phase.  When the heat release events 
occur,  we want to run the simulation at the finest grid granularity possible, 
and store the data as frequently as possible.  Therefore, it is imperative to 
be able to predict the start of the heat release event using an indicator and 
trigger that serve to inform the application to adjust its grid resolution and 
I/O frequency accordingly, see Figure~\ref{fig:heat_release}. 

\section{Designing a Noise-Resistant \\Indicator and Trigger} 
\label{sec:indicator_trigger}
In this section we describe the design of an indicator and trigger for heat release for our combustion use case.   To provide context, we begin by discussing the intuitions that informed our design.

\subsection{Chemical Explosive Mode Analysis}
\label{sec:intution}
Chemical explosive mode analysis (CEMA) is known to be a reliable technique to predict incipient heat release. 
Here  we  provide brief description and refer to  \cite{lu,shan} for details. 
The conservation equations for reacting species can be written as $\frac{D\mathbf{y}}{Dt} = \mathbf{\omega(y)} + \mathbf{s(y)}$.
\noindent The vector $\mathbf{y}$ represents temperature and reacting
species mass fractions, $\mathbf{\omega}$ is the reaction source term and
$\mathbf{s}$ is the mixing term. Then, the Jacobian is  $\mathbf{J_\omega} +
\mathbf{J_s}$ where
$\mathbf{J_\omega} =  \frac{\partial \mathbf{\omega(y)}}{\partial \mathbf{y}}$ and
$\mathbf{J_s}  =  \frac{\partial \mathbf{s(y)}}{\partial \mathbf{y}}$. 
The eigen-decomposition of the chemical Jacobian, $\mathbf{J_\omega}$ can be used to infer
chemical properties of the mixture.  
 Let $\lambda_e$ be the eigenvalue with the largest real part. 
$\lambda_e$ is defined as a chemical explosive mode (CEM) if 
$\textrm{Re}(\lambda_e) > 0$, which indicates that point will undergo ignition. If it has undergone ignition, we
have $Re(\lambda_e)<0$.   The presence of a CEM indicates the propensity of a mixture to ignite.

\begin{figure}[ht]
\centering
\includegraphics[width=1.0\columnwidth ]{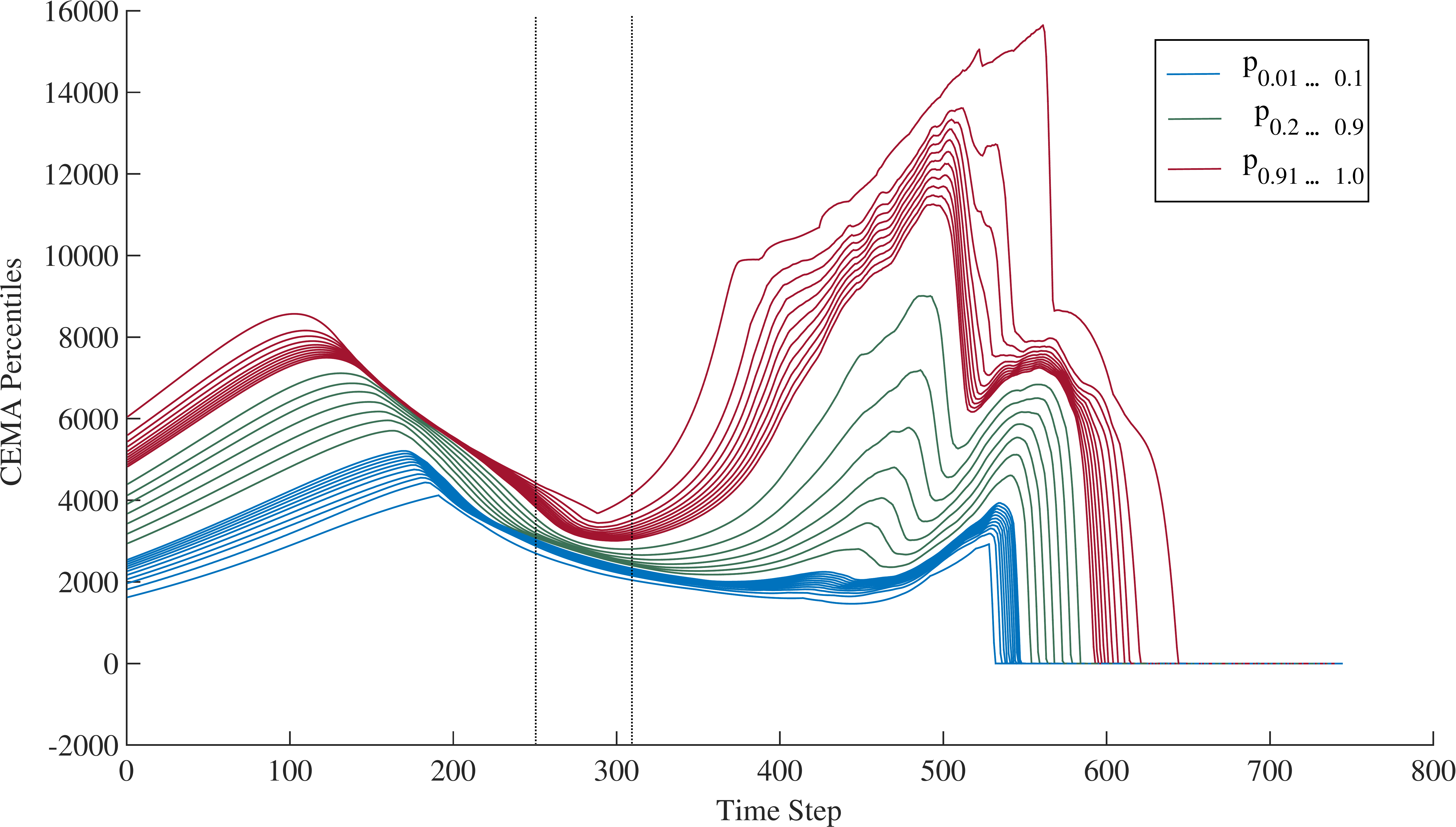}
\caption{\label{fig:CEMA_indicator} Percentile plot of CEMA values. 
The blue curves correspond to $p_{0.01, 0,02, \ldots , 0.1}$, the green curves to $p_{0.2, 0.3, \ldots ,0.9}$ and the red
curves to $p_{0.91, 0.92,  \ldots ,1}$. }
\end{figure}

Our CEMA-based indicator is based on global trends of CEMA over time.
Consider Figure~\ref{fig:CEMA_indicator} which provides a summary of the trends of CEMA
values across all time steps in a simulation. At timestep $t$, let $\cema(t)$ be
the sorted (in nondecreasing order) array of CEMA values on the underlying mesh. 
For $\alpha \in (0,1]$, the $\alpha$-percentile is the entry
$\cema(t)_{\lceil \alpha N \rceil}$. More specifically, it is the
value in $\cema(t)$ that is greater than at least $\cei{\alpha N}$
values in $\cema(t)$. We denote this value by $\pt_\alpha(t)$.

We notice that as the simulation progresses, the distance between the higher
percentiles (red curves) decreases then suddenly increases. This is illustrated
in the plot by the spread in the red curves that occurs between the dashed
lines defined by the domain expert that indicate the desired transition phase.  Our goal is to capture this empirical observation with a metric,  and use it to predict  heat release.
 
Our empirical observation is consistent with the underlying physics, and we refer to~\cite{cema-insitu} for a detailed discussion on  the underlying physics. 

\subsection{Designing a Noise-Resistant CEMA-Based Indicator}
\label{sec:indicator}
We introduce an indicator function, \emph{\cmetric-indicator} that quantifies
the distribution of the top quantiles of CEMA values over time. The
\cmetric-indicator is equal to the coefficient of variation (COV) of the top
percentiles of CEMA values, i.e. the ratio of their standard deviation to
their mean. 

Formally, quantiles are defined as  values taken at regular intervals from the
inverse of the cumulative distribution function of a random variable. For a
given data set, quantiles are used to divide the data into equal sized sets 
after sorting, and the quantiles are the values on the boundary between
consecutive subsets.  A special case is dividing into 100 equal groups, when we can
refer to quantiles as percentiles.  This paper focuses on percentiles with
numbers in the $[0,1]$ range (although all techniques presented here can be
generalized for any quantiles). For example, the $0.5$ percentile will refer to
the median of the data set. 
 
We define our indicator using 2 parameters: $\alpha$ and
$\beta$.  
We  want to detect whether the range
covered by top quantiles shrinks, and $\alpha$ represents the lower end of the
top percentiles, whereas $\beta$ represents the top percentile considered. Therefore, the  range of top percentiles we  measure is
$\left[\pt_\alpha(t), \pt_\beta(t)\right]$.  In our indicator, we choose $\alpha
< \beta$ (typically in the range $[0.90,0.99]$). We measure the spread at time
$t$ by the COV of the $\pt_s(t)$ values for $\alpha \le s \le \beta$ which
results in the following \cmetric-indicator:
\be
\cmetric_{\alpha,\beta}(t) = \sqrt{\frac{\mu}{N-1} \sum_{s=\alpha}^{\beta}
(p_s/\mu - 1)^2}
\label{eq:cmetric}
\ee

\noindent where
\be
\mu = \frac{1}{N} \sum_{s=\alpha}^{\beta} p_s
\label{eq:mu}
\ee

\noindent and $N$ is the number of percentile values considered between $\alpha$ and
$\beta$. For instance, if $\alpha=0.90$ and $\beta=0.99$, we would obtain
$N=10$.

\begin{figure}[ht]
\begin{center}
\hspace*{-5ex}\includegraphics[width=1.2\columnwidth]{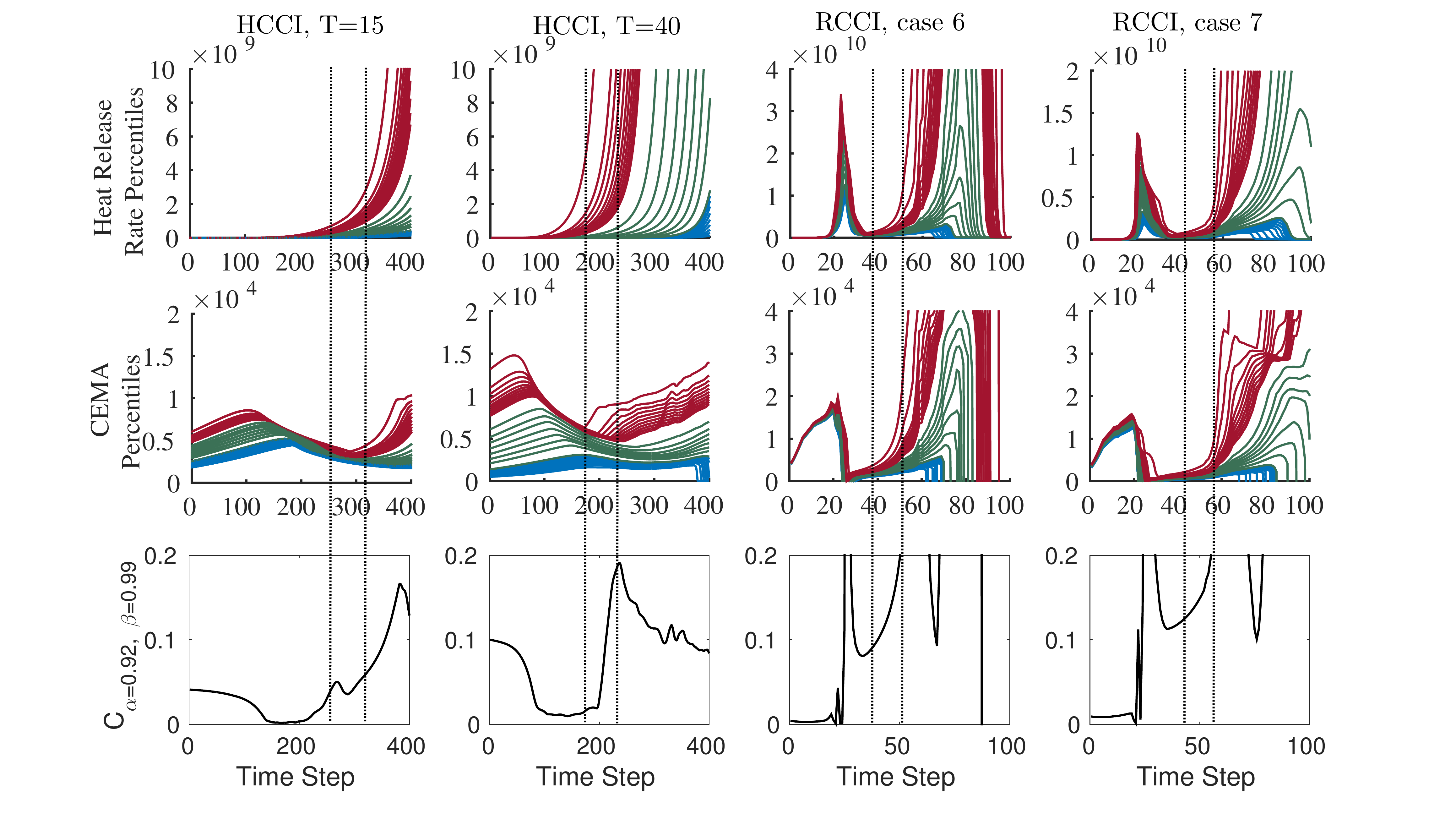}
\caption{\label{fig:CEMA} Plots showing the (top row) percentiles of the
heat release rate, (middle row) percentiles of CEMA, and (bottom row) the  \cmetric-indicator, as indicated.  The blue curves correspond to $p_{0.01, 0.02, \ldots , 0.1}$, the green curves to $p_{0.2, 0.3,  \ldots , 0.9}$ and the red curves to $p_{0.91, 0.92,  \ldots , 1}$. The \cmetric-indicator shown is
evaluated for $\alpha_0=0.92$ and $\beta=0.99$. The vertical dotted lines crossing the images indicate a window of
acceptable ``true'' trigger time steps, as identified by a domain expert.
For the RCCI cases, the trigger time ranges are based on the High Temperature
Heat Release (HTHR), i.e., the second peak in the Heat Release Rate (HRR) profiles.
}
\end{center}
\end{figure}

Figure~\ref{fig:CEMA} illustrates percentile plots for heat release (top row) and
CEMA
(middle row).  In the percentile plots, the lowest
blue curve and the highest red curve correspond to the 1 and
100 percentiles ($p_{0.01}$ and $p_{1}$), respectively. The blue curves
correspond to $p_{0.01, 0.02, \ldots, 0.1}$, the green curves to $p_{0.2, 0.3,  \ldots , 0.9}$
and the red curves to $p_{0.91, 0.92,  \ldots, 1}$. 
The \cmetric-indicator evaluated for $\alpha=0.92$ and
$\beta=0.99$  is shown in the bottom row of Figure~\ref{fig:CEMA}. 
Results are generated for four test cases described in Table~\ref{tab:usecases}.
The vertical dotted lines were identified  by a domain expert who, via examination
of heat release and CEMA percentile plots, visually
located the range of time steps in the simulation where the mesh resolution and I/O
frequency should  be increased. We refer to this range of time steps  
as the ``true'' trigger range we wish to identify with the 
\cmetric-indicator and trigger functions. Note, for the RCCI cases, there are
two ignition ranges.
To simplify the following exposition, we focus on the second rise in the heat release rate
profiles, as this is the ignition stage of interest to the scientists. However,
we note that our approach is robust in identifying the first ignition stage as
well.

\begin{table*}[tb]
{\small
  \caption{\label{tab:usecases} Four Combustion Use Cases analyzed in this
study. The ``true'' trigger time ranges are estimated based on $95-100^{\textrm{th}}$
percentiles of the heat release rate. The computed time ranges were evaluated
using our quantile sampling approach.  The \cmetric-trigger and \pmetric-trigger were
computed over 50 realizations of experiments with 20 samples per process.
}
\begin{center}
\begin{tabular}{|c|c|c|c|c|c|c|} 
 \hline
 Problem & Number of & Number of & ``True'' Trigger  & \cmetric-trigger &
 \pmetric -trigger  & Total \\
 Instance & Grid Points & Species & Time Range & Detection &
 Detection~\cite{cema-insitu}  & Processes \\ \hline\hline
 HCCI, T=15   & 451,584     & 28  & 250-315 & 235 - 265  & 250 - 262 & 1600 \\
 \hline
 HCCI, T=40   & 451,584     & 28  & 175-225 & 180 - 220  & 213 - 220 & 1600 \\
 \hline
 RCCI, case 6 & 2,560K   & 116 & 38-50   & 34 - 38 & 28 - 45 &  6400 \\
 \hline
 RCCI, case 7 & 2,560K--10,240K   & 116 & 42-58   & 32.5 - 35.5 & 35 - 50 & 6400 \\
 \hline
 \end{tabular}
 \end{center}
 } 
 \end{table*}

\subsection{Defining a Trigger}
\label{sec:trigger} 

In addition to defining a noise-resistant indicator function, we also need to define a trigger
function that returns a boolean value, capturing whether a property of the
indicator has been met. Looking at Figure~\ref{fig:CEMA}, we notice that across all
experiments from Table~\ref{tab:usecases}, the \cmetric-indicator is increasing during
the true trigger time step windows.  Therefore, we seek to find a value 
$\thresh_{\cmetric} \in (0,1)$, such that $\cmetric_{\alpha,\beta}(t)$
crosses $\thresh_{\cmetric}$ \emph{from
below}, as the simulation time $t$ progresses. 

\begin{figure}[t]
\centering
\includegraphics[width=0.7\columnwidth]{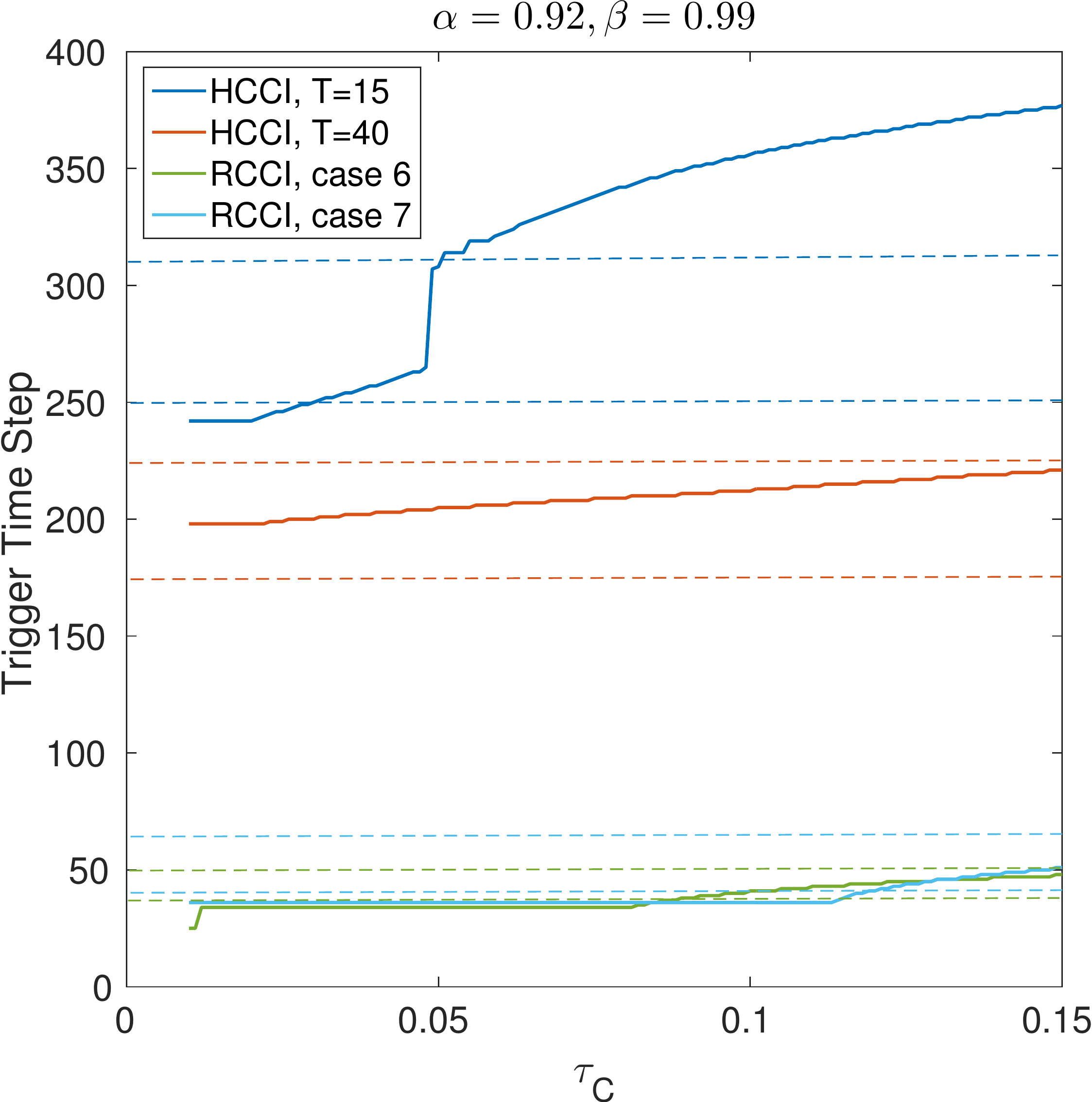}
\caption{\label{fig:Triggers_thresh} This figure plots the trigger time steps as a function of
$\thresh_{\cmetric}$ for $\cmetric_{\alpha=0.92, \beta=0.99}(t)$.
There is a range of viable values of $\thresh_{\cmetric} \in [0.0, 0.05]$
that predict early stage heat release.
}
\end{figure}

Figure \ref{fig:Triggers_thresh} plots the trigger time steps as a function of
$\thresh_{\cmetric}$ for $\cmetric_{\alpha=0.92, \beta=0.99}(t)$. 
This plot shows that the viable range of $\thresh_{\cmetric}$ is $[ 0.01, 0.05]$
for the four use cases described in Table~\ref{tab:usecases}.  
The horizontal dashed lines indicate the true trigger range identified by our
domain expert.  We consider those values of $\thresh_{\cmetric}$ that fall
within the horizontal dashed lines to be good $\thresh_{\cmetric}$ values for our
trigger, with those values of $\thresh_{\cmetric}$ falling below the dashed
lines still considered viable.  
 We note the plots for other values of
$\alpha$ and $\beta$ look similar and have been omitted due to lack of space in
this text. 

\section{Computing Indicators and \\Triggers Efficiently: \\A Sublinear Approach}
\label{sec:sample} 
The previous section showed that \cmetric-indicator and
trigger are robust to noise fluctuations and act as a precursor to rapid heat
release in combustion simulations. In~\cite{cema-insitu} we show 
the computational cost of computing CEMA-based indicators can be prohibitive for large-scale
simulations (up to 60 times the cost of a simulation time-step depending on the
simulation parameters).  
To mitigate the cost, we introduced a quantile sampling approach that comes with
provable bounds on accuracy as a function of the number of samples.  Most
importantly,  the required number of samples for a specified accuracy is
independent of  the size of the problem, hence our sampling based algorithms
offer  excellent scalability. 

We summarize the quantile sampling method here and refer the reader
to~\cite{cema-insitu} for analytical and empirical proofs of convergence. Consider an array $A \in \RR^N$, in sorted order.
Our aim is to estimate the $\alpha$-percentile of $A$. (We use $\pt_\alpha$ to denote
the percentiles.) Note that this 
is exactly the entry $A_{\lceil \alpha N \rceil}$. Here is a simple sampling procedure. \\[-1ex]
\begin{compactenum}
	\item Sample $\samp$ independent, uniform indices $r_1, r_2, \ldots, r_\samp$
	in $\{1,2,\ldots,N\}$. \\ Denote by $\widehat{A}$ the sorted array $[A(r_1),A(r_2),\ldots,A(r_\samp)]$.
	\item Output the $\alpha$-percentile of $\widehat{A}$ as the estimate, $\widehat{\pt}_\alpha$. \\[-1ex]
\end{compactenum}

\begin{figure}[t]
\centering
\includegraphics[width=1.\columnwidth]{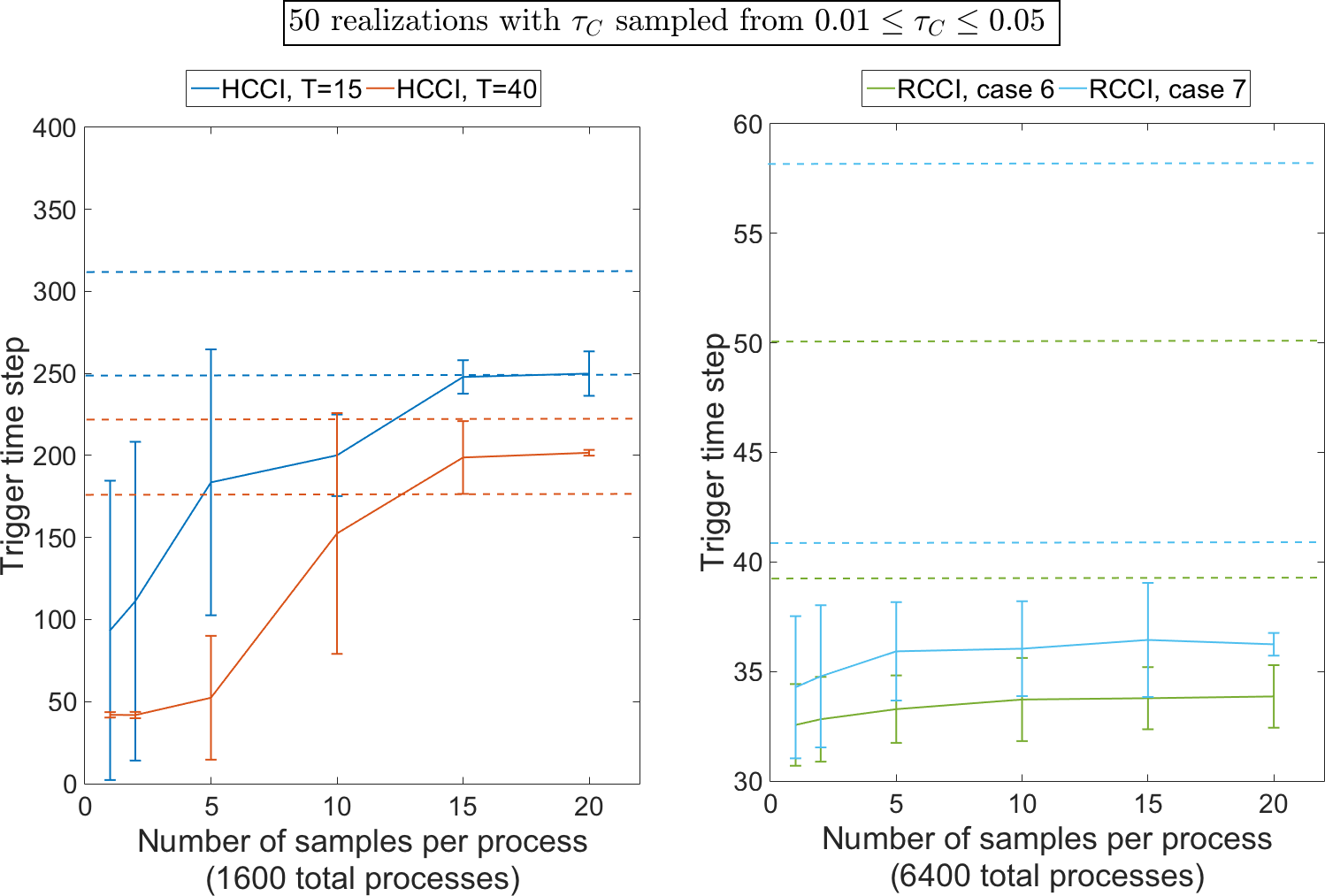}
\caption{\label{fig:sampling} Plots illustrating the variability of the trigger
time steps 
predicted by the \cmetric-indicator and trigger as a function of the number of
samples per process.  The data for these plots was generated via 50
realizations of the \cmetric-indicator with $\alpha=0.92$ and $\beta=0.99$, and
$\thresh_{\cmetric}$ drawn from $[0.01, 0.05]$.The horizontal dashed lines
define the range of time steps within which we would like to make the workflow
transition (as identified by a domain expert.}
\end{figure}

We performed a series of experiments examining the variation in the trigger time steps 
as a function of the number of samples used per process.  The data for
Figure~\ref{fig:sampling} was generated via 50 realizations of  the
\cmetric-indicator with $\alpha=0.92$ and $\beta=0.99$,
with $\thresh_{\cmetric}$ drawn from $[0.01, 0.05]$. The
horizontal dashed lines in this figure define the range of true trigger time steps within
which we would like to make the workflow transition (as identified by a domain expert).
This plot demonstrates that, even across the range of $\thresh_{\cmetric}$
values,  with a small number of samples per process, our quantile sampling
approach can accurately estimate the  true trigger time steps as defined by the
domain expert. 
The accuracy of our quantile-based sampling predictions is also shown in Table~\ref{tab:usecases}, which lists the the triggers predicted by the \cmetric-indicator and
\pmetric-indicator that we used in our previous work~\cite{cema-insitu}. The \pmetric-indicator is computed as $\pmetric_{\alpha,\beta,\gamma} = \frac{\pt_\alpha - \pt_\gamma}{\pt_\beta - \pt_\gamma}$,
for $\alpha=0.94$, $\beta=0.98$, $\gamma= 0.01$,  and the trigger threshold chosen as $[0.725, 0.885]$. The results show both methods are accurate. We are currently investigating  how  we can improve our results by using two metrics concurrently. Note that  using both metrics will not induce an additional burden, since the two metrics  can be computed from the same set of samples.  

\section{Conclusion}
\label{sec:conc} 
We propose a new indicator and trigger 
for making data-driven control-flow decisions in~situ. Using a provably robust
sampling-based approach, 
of CEMA values (which, when computed in full, can cost up to 60 times a
simulation time step).  Our experiments show that our proposed indicator, based
on the coefficient of variation, can  efficiently predict rapid increases in heat
release.  We believe the new metric 
can be deployed collectively with a previously defined mechanism, 
to  provide more robust detections, since together they will be  less prone to false positives. We  note that  using both metrics 
will not induce an additional burden, since the two metrics  can be computed
from the same set of samples.  Our future work aims to explore deployment of
both metrics jointly in production simulation runs.

\end{document}